\documentclass[conference]{IEEEtran}
\IEEEoverridecommandlockouts
\usepackage{cite}
\usepackage{amsmath,amssymb,amsfonts}
\usepackage{algorithmic}
\usepackage{graphicx}
\usepackage{textcomp}
\usepackage{xcolor}
\usepackage[linkcolor=black,colorlinks=true,citecolor=black,filecolor=black]{hyperref}
\def\BibTeX{{\rm B\kern-.05em{\sc i\kern-.025em b}\kern-.08em
    T\kern-.1667em\lower.7ex\hbox{E}\kern-.125emX}}
\usepackage[utf8]{inputenc}

\usepackage{dirtytalk}
\usepackage{subcaption}
\usepackage{tablefootnote}

\IEEEoverridecommandlockouts
\IEEEpubid{\makebox[\columnwidth]{978-0-7381-2384-4/20/\$31.00~\copyright{}2020 IEEE \hfill} \hspace{\columnsep}\makebox[\columnwidth]{ }}

\begin{document}

\title{Performance of AV1 Real-Time Mode}
\author{\IEEEauthorblockN{1\textsuperscript{st} Ludovic Roux}
\IEEEauthorblockA{\textit{CoSMo Software}\\
Singapore \\
ludovic.roux@cosmosoftware.io}
\and
\IEEEauthorblockN{2\textsuperscript{nd} Alexandre Gouaillard}
\IEEEauthorblockA{\textit{CoSMo Software}\\
Singapore \\
alex.gouaillard@cosmosoftware.io}
}
\date{August 2020}

\maketitle

\begin{abstract}
With COVID-19, the interest for digital interactions has raised, putting in turn real-time (or low-latency) codecs into a new light. Most of the codec research has been traditionally focusing on coding efficiency, while very little literature exist on real-time codecs.
It is shown how the speed at which content is made available impacts both latency and throughput. The authors introduce a new test set up, integrating a paced reader, which allows to run codec in the same condition as real-time media capture.
Quality measurements using VMAF, as well as multiple speed measurements are made on encoding of HD and full HD video sequences, both at 25~fps and 50~fps to compare the respective performances of several implementations of the H.264, H.265, VP8, VP9 and AV1 codecs.
\end{abstract}


\section{Introduction}

\subsection{Codec Use Cases}
The term codec usually refers to algorithms that encode, respectively decode, binary representation of media.

Currently there are arguably three major use cases for large-scale codec usage:
\begin{itemize}
    \item encoding of original raw content, client side,
    \item trans-coding of content, server-side,
    \item decoding of content on the receiving side.
\end{itemize}

Codecs are usually evaluated based on coding efficiency \cite{decock2016} \cite{akyazi2018}, a.k.a compression/quality ratio achieved against run time. Comparisons are made using the Bj{\o}ntegaard rate difference (BD-rate) \cite{bjontegaard2001} \cite{bjontegaard2008} and multiple papers exist to guide researchers choosing the most representative dataset, the most meaningful metric, and the best representation of results \cite{ietf-netvc2020}.

For example, the Xiph.org Foundation has developed the very extensive AreWeCompressedYet~\cite{arewecompressedyet} automatic service to enable comparisons between different implementations of video codecs using various metrics.

With COVID-19 and the new normal, people have a new appetite for more interactive uses of streaming, to get the same experience and value they were enjoying in real-life.
Correspondingly, it has increased the demand for faster-than-live streaming, ``live'' being 5 seconds behind real-time, to reach the less-than 500~ms. level of latency where human interactions thrive. In parallel, the rise of cloud gaming, augmented reality (AR) and virtual reality (VR) have been pushing video codecs in the same direction, albeit with generated content instead of natural content and even higher expectation for latency (150~ms. max) \cite{stadia105}.

In this paper we focus on the specific use case of real-time consumption of media for interactive applications. This use case puts a specific emphasis on latency, making coding efficiency a secondary target.

\subsection{Pre-Recorded Content Encoding}

The encoding \emph{throughput} is the total number of frames of the input divided by the total duration of the encoding process. 
The encoding \emph{latency} is the time it takes for a single frame to go through the encoding process.

When the content is readily available, the encoding process can be distributed, reducing the encoding time. The total run time used in coding efficiency computation includes both latency and processing time, diluting out latency. Latency should be reported separately to be comparable.

In Video-on-Demand (VOD), latency is usually negligible for both encoder and decoder compared to the encoding time. Latency is more than often not taken into account, and the main parameter of the encoder will be the speed (1 to 8), which represent a coding efficiency vs complexity compromise. Implicitly, the quality follows the speed, although not linearly.



As such, the delay or latency induced by the encoding and decoding processes are almost never taken in account. Actually, assuming latency is negligeable, and the entire media is available, many encoders use a 2-pass approach to gain more efficiency per file/movie. Netflix is known to pioneer distributed ``per-chunks,'' ``per-shots'' techniques investing time and resources in the encoding process in exchange for maximum efficiency \cite{netflix-2018}.

The most recent codecs performance study~\cite{chen2020} compared libaom AV1 encoder against x265 and libvpx-vp9 using their best quality mode and a two-pass compression.

\subsection{ Live Content Capture Speed }

A big difference between recorded content and real-time content is the speed at which frames become available and corresponding impact on latency. For example, real-time or live media needs to be encoded in one-pass and not in two or more passes like algorithms focused on coding efficiency would.

\subsubsection{Real-time encoder cannot encode faster than real-time}

Even if the encoder was able to encode one frame at a time, at a faster-than-real-time speed, you still need to wait on the capturer to provide the next frame before the encoder can process it. 

\subsubsection{Real-time encoder latency is correlated with buffer depth}

Let's take the use case where you have a 60 frames deep frame buffer. With live content, you need to wait until those frames are generated before you can start any kind of processing. At a capture rate of 30 fps, one must wait 2~seconds before the above buffer is ready to be used, even if the encoder can then encode faster than 30 fps.

To be able to compare traditional codec (default settings), and real-time modes of certain codecs implementation under the same condition, we need to make sure that the frames are being delivered to the encoder at real-time speed even if it is pre-recorded content. 

\subsection{Codecs Performance in Real-Time Use Cases}

The run time is latency plus encoding time, and the latency is a function of both the depth of any frame buffer and the constant capture rate. Increasing the frame rate reduces latency but increases the work load. The easiest way to reduce the latency is to reduce the size of the frame buffer, or to remove the need for a frame buffer altogether.

This is not to be confused with encoders speed setting. Most recent codecs design involves sub algorithms, referred to as ``tools.'' Some tools are more demanding in term of complexity, or latency than others, and not all have the same impact on efficiency.  As described in \cite{yu2003} figure 1, the motion estimation is the dominant contributor to run time budget in encoding. 

The speed settings maps to the choice of some specific subset defining a certain trade off between coding efficiency and encoding speed (possibly disregarding completely the latency). Some other codecs also have an explicit real-time mode, settings that relate to coding/efficiency -- latency trade off.

This comes at a cost in terms of coding efficiency, and makes the real-time mode of codecs not directly comparable with  rate-compression graphs. There is no study to date about how much coding efficiency you trade off for decreased latency.

In this paper, we will concentrate on the real-time mode of some of the encoder implementations for H.264 (AVC), VP8, VP9, and AV1 which are all used in the webrtc.org code today. An implementation of H.265 (HEVC) will be also used for comparison.

\begin{figure}[tb]
    \centering
    \begin{subfigure}[b]{0.24\textwidth}
        \centering
        \includegraphics[width=\textwidth]{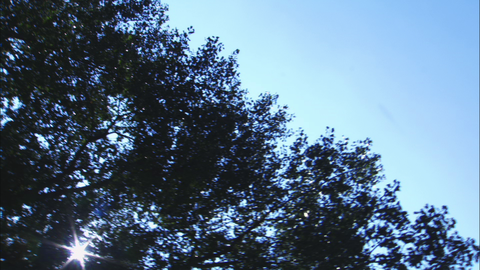}
        \caption{Blue sky (BS25)}
    \end{subfigure}%
    ~ 
    \begin{subfigure}[b]{0.24\textwidth}
        \centering
        \includegraphics[width=\textwidth]{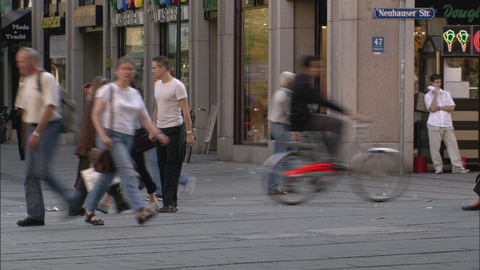}
        \caption{Pedestrian area (PA25)}
    \end{subfigure}
    \newline
    \begin{subfigure}[b]{0.24\textwidth}
        \centering
        \includegraphics[width=\textwidth]{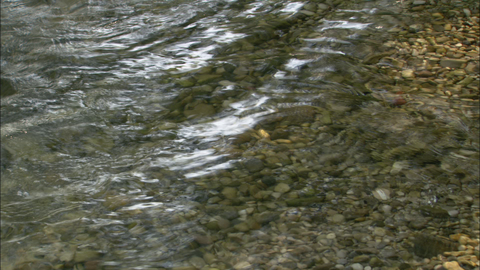}
        \caption{Riverbed (RB25)}
    \end{subfigure}%
    ~
    \begin{subfigure}[b]{0.24\textwidth}
        \centering
        \includegraphics[width=\textwidth]{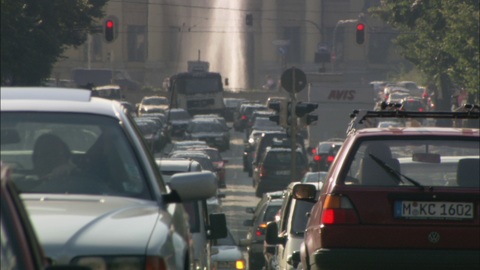}
        \caption{Rush hour (RH25)}
    \end{subfigure}
    \newline
    \begin{subfigure}[b]{0.24\textwidth}
        \centering
        \includegraphics[width=\textwidth]{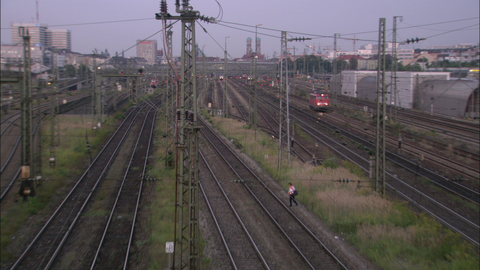}
        \caption{Station2 (ST25)}
    \end{subfigure}%
    ~
    \begin{subfigure}[b]{0.24\textwidth}
        \centering
        \includegraphics[width=\textwidth]{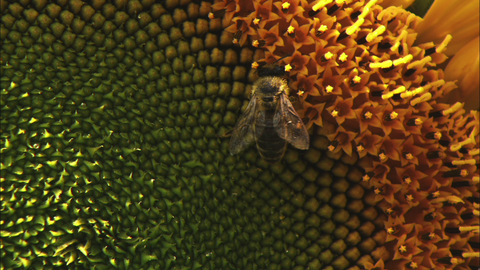}
        \caption{Sunflower (SF25)}
    \end{subfigure}
    \newline
    \begin{subfigure}[b]{0.24\textwidth}
        \centering
        \includegraphics[width=\textwidth]{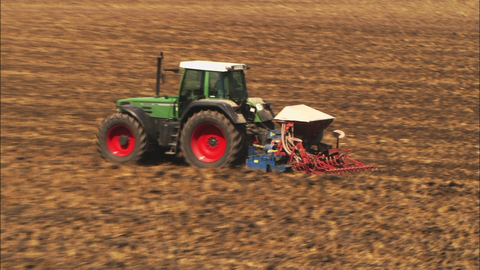}
        \caption{Tractor (TR25)}
    \end{subfigure}%
    ~
    \begin{subfigure}[b]{0.24\textwidth}
        \centering
        \includegraphics[width=\textwidth]{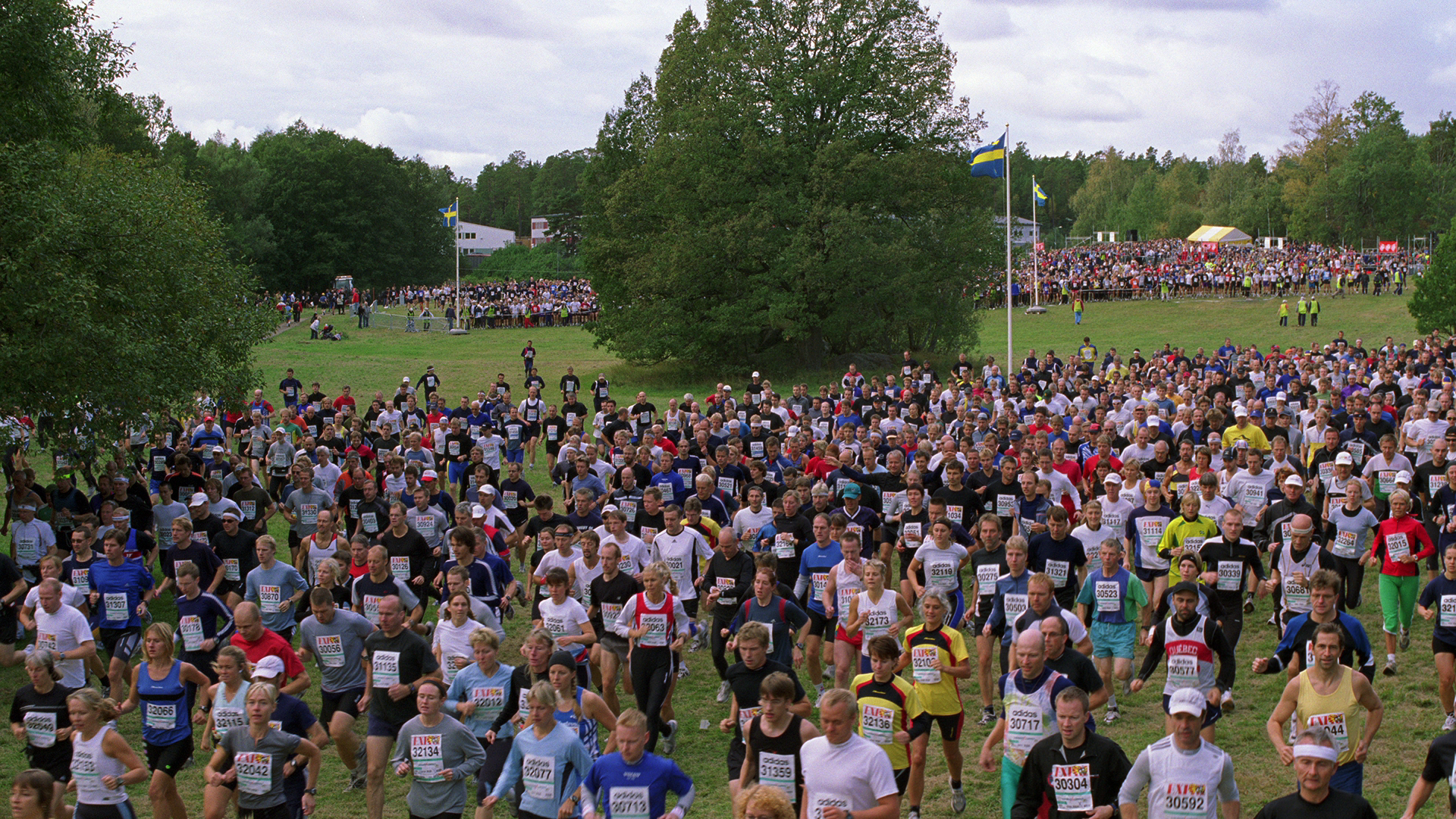}
        \caption{Crowd run (CR50)}
    \end{subfigure}
    \newline
    \begin{subfigure}[b]{0.24\textwidth}
        \centering
        \includegraphics[width=\textwidth]{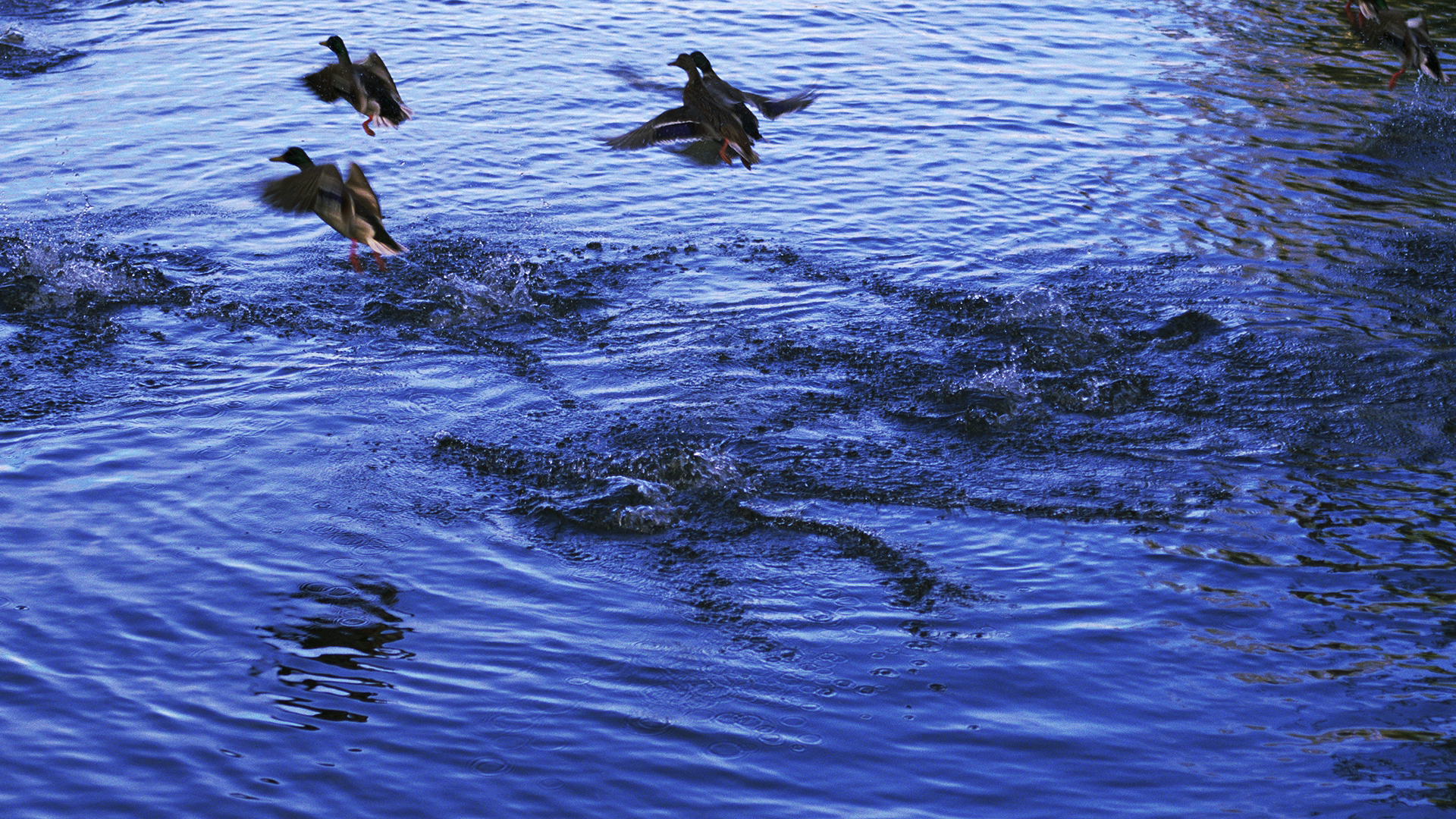}
        \caption{Ducks take off (DT50)}
    \end{subfigure}%
    ~
    \begin{subfigure}[b]{0.24\textwidth}
        \centering
        \includegraphics[width=\textwidth]{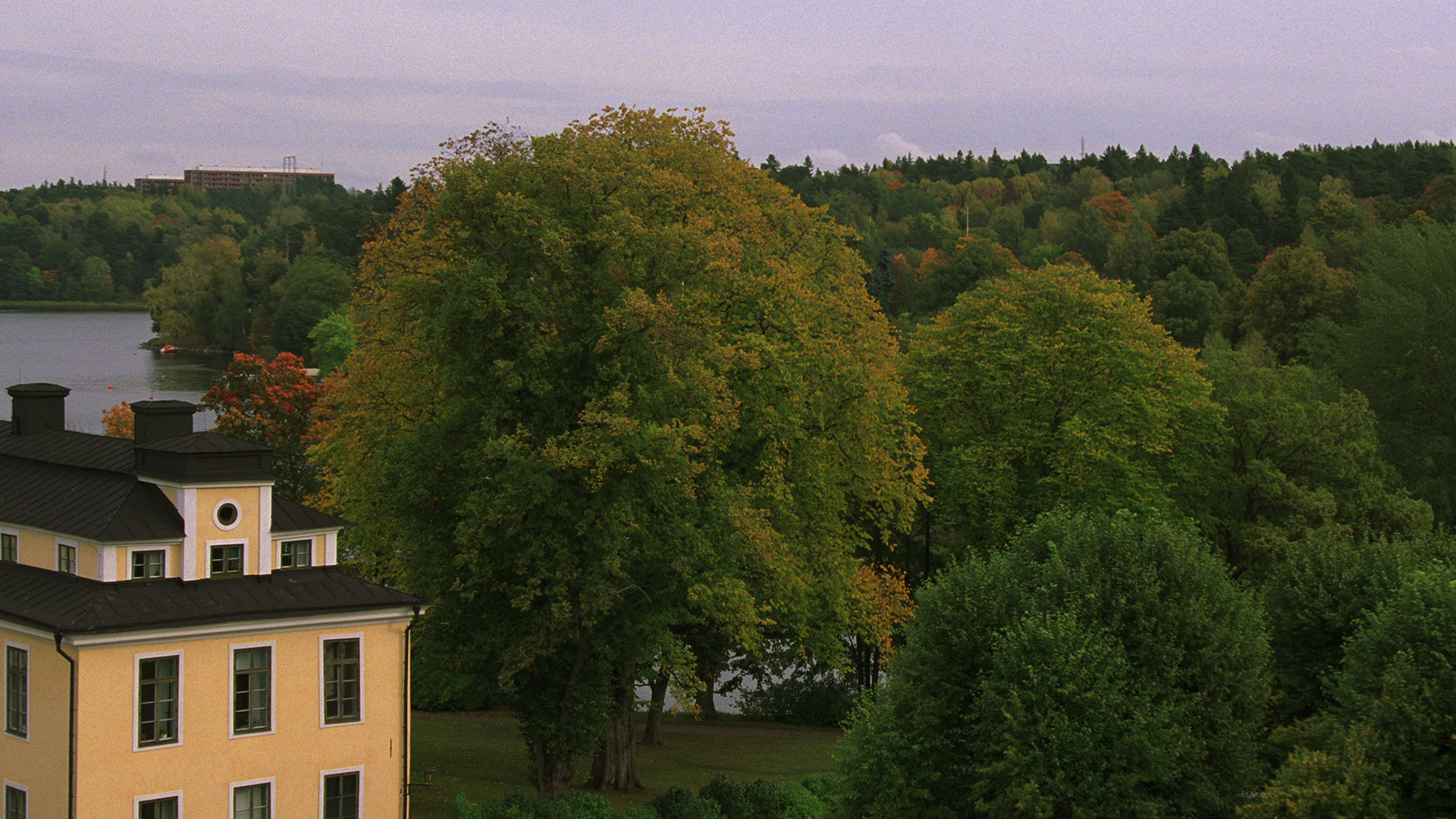}
        \caption{In to tree (IT50)}
    \end{subfigure}
    \newline
    \begin{subfigure}[b]{0.24\textwidth}
        \centering
        \includegraphics[width=\textwidth]{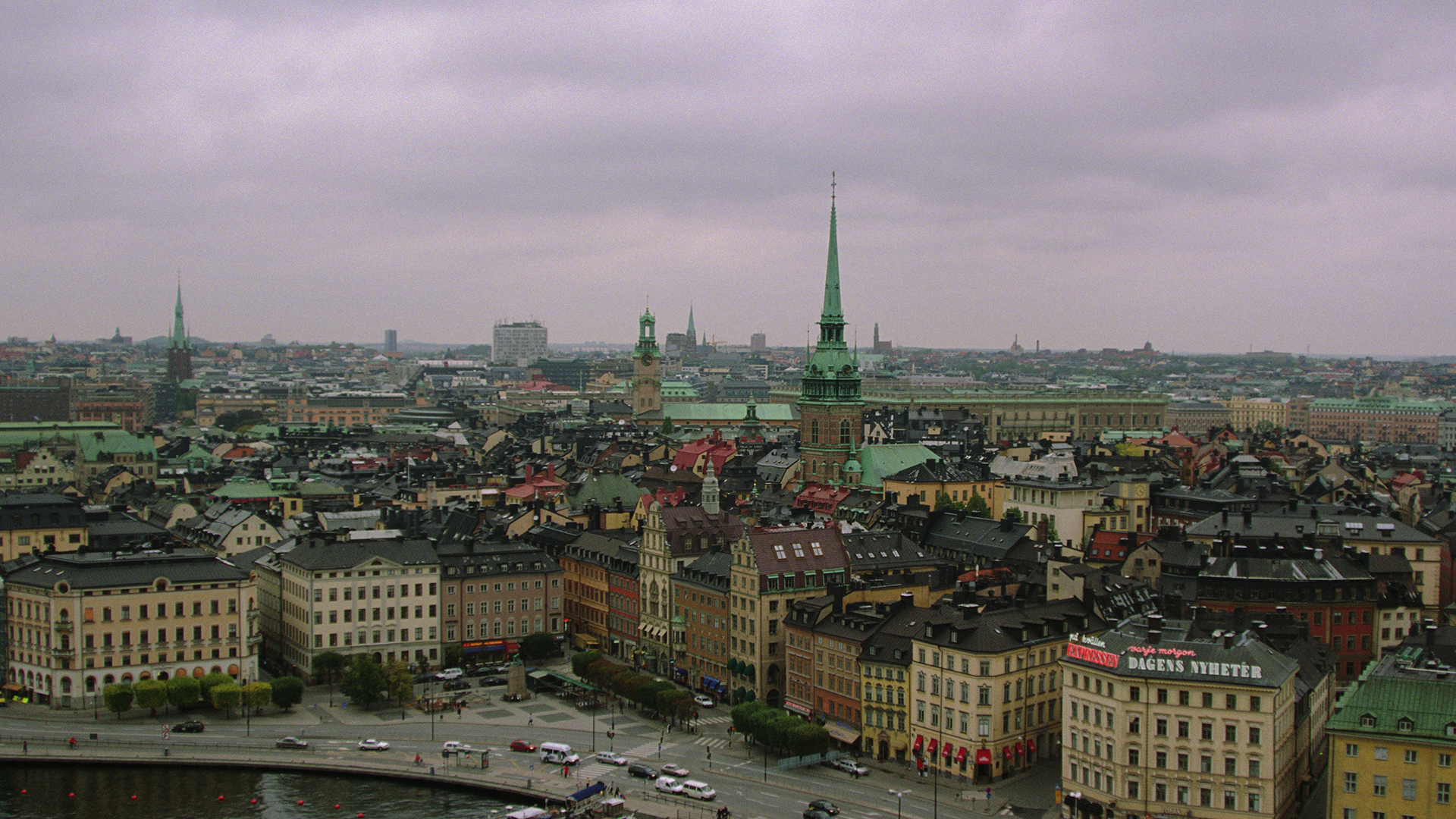}
        \caption{Old town cross (OT50)}
    \end{subfigure}%
    ~
    \begin{subfigure}[b]{0.24\textwidth}
        \centering
        \includegraphics[width=\textwidth]{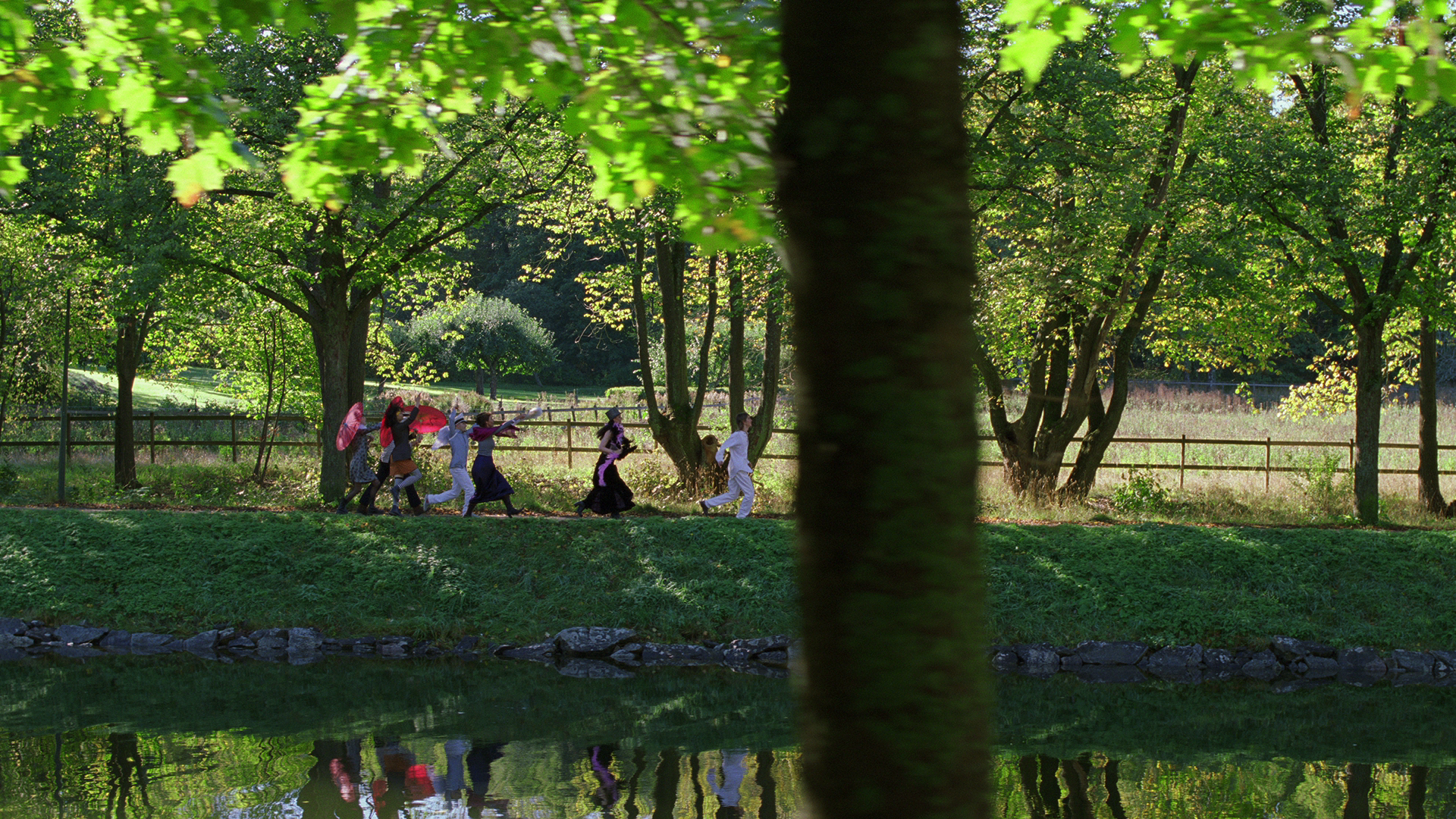}
        \caption{Park joy (PJ50)}
    \end{subfigure}
    \caption{Snapshots of each video sequences}
    \label{fig:snapshots}
\end{figure}

\begin{table}[tb]
    \centering
    \begin{tabular}{c|c|c|c}
        VMAF & MOS & ACR label  & Error visibility \\
        \hline
        100  &  5  & excellent  & Imperceptible \\
         80  &  4  & good       & Perceptible error, not annoying \\
         60  &  3  & fair       & Visible error, slightly annoying \\
         40  &  2  & poor       & Visible error, annoying \\
         20  &  1  & bad        & Visible error, very annoying
    \end{tabular}
    \caption{Mapping of VMAF scores to MOS values}
    \label{tab:VMAF-MOS}
\end{table}

\begin{table*}[tb]
    \centering
    \begin{tabular}{c|c|c|c|c|c|c}
        Filename        & Short name & fps & Duration & \# frames & Content type                      & Camera motion \\
        \hline
        Blue sky        & BS25      & 25  & 8.68 s.  & 217       & trees and sky                     & slow motion \\
        Pedestrian area & PA25      & 25  & 15 s.    & 375       & people walking                    & static \\
        Riverbed        & RB25      & 25  & 10 s.    & 250       & light reflection on waves, riverbed visible & static \\
        Rush hour       & RH25      & 25  & 20 s.    & 500       & cars in traffic                   & static \\
        Station2        & ST25      & 25  & 12.52 s. & 313       & railway tracks and train          & zoom out \\
        Sunflower       & SF25      & 25  & 20 s.    & 500       & closeup on bee foraging sunflower & slow motion \\
        Tractor         & TR25      & 25  & 27.6 s.  & 690       & tractor plowing a field           & slow motion tracking, zoom in and zoom out \\
        \hline
        Crowd run       & CR50      & 50  & 10 s.    & 500       & people running                    & slow motion \\
        Ducks take off  & DT50      & 50  & 10 s.    & 500       & ducks taking off, waves on a lake & static \\
        In to tree      & IT50      & 50  & 10 s.    & 500       & park, closeup on tree             & zoom in to a tree \\
        Old town cross  & OT50      & 50  & 10 s.    & 500       & city old centre                   & slow motion \\
        Park joy        & PJ50      & 50  & 10 s.    & 500       & people running along a canal      & motion tracking
    \end{tabular}
    \caption{List of 12 video sequences used in this study}
    \label{tab:dataset}
\end{table*}

\section{Methodology}

\subsection{Quality Metric}

Compression can be objectively evaluated. Quality of video frames can also be evaluated objectively, however it has been shown that scores provided by objective metrics like PNSR (Peak Signal to Noise Ratio) \cite{ANSI2001} poorly correlate with human evaluation of image or video quality \cite{sheikh2006}. To address this drawback, subjective metrics like VMAF (Video  Multimethod  Assessment  Fusion) \cite{vmaf2016,vmaf_github} were introduced. A recent study on the evaluation of objective video quality metrics has demonstrated a good correlation between subjective scores given by humans and VMAF scores~\cite{lee2017}. We have followed the latest trend in codec research and used VMAF in this study.

To give an interpretation of a VMAF score, one can relate it to the typical Mean Opinion Score (MOS) value ranging from 1 to 5. A very common rating scale for MOS is the Absolute Category Rating (ACR) methodology~\cite{itu2008_p910}: ``bad,'' ``poor,'' ``fair,'' ``good'' and ``excellent.'' VMAF gives a score in the range $[0,100]$. VMAF score 20 can be mapped to ``bad,'' score 40 to ``poor,'' score 60 to ``fair,'' score 80 to ``good,'' and score 100 to ``excellent''~\cite{vmaf2018}. Table~\ref{tab:VMAF-MOS} gives a synthetic relationship between VMAF scores and MOS values.

\subsection{Datasets}

For easier comparison of results, we have used video sequences having the same resolution, the same color space and the same bit depth. Only the duration or the frame rate of video sequences is different from one video to another.

We have focused on 1080p HD video. This is the resolution recommended to compute VMAF scores using the default model v0.6.1 \cite{vmaf2018}. Table~\ref{tab:dataset} gives the list of the 12~videos used in our study. 

All the video sequences use YUV format, 8~bits depth and are not compressed. They have been selected from the publicly available Xiph.org Video Test Media [derf's collection] dataset.\footnote{Xiph.org Video Test Media [derf's collection] \url{https://media.xiph.org/video/derf/}}

There are two groups of videos: a group of 7~videos having a frame rate of 25~fps, and a group of 5~videos with a frame rate of 50~fps. Fig.~\ref{fig:snapshots} shows a snapshot of each video.

\begin{table*}[tb]
    \centering
    \begin{tabular}{c|c|c|p{7cm}}
        Codec    & Encoder and version & Source code                                    & Configuration options \\
        \hline
        AV1      & libaom 2.0.0   & \url{https://aomedia.googlesource.com/aom/}         & \texttt{cmake -DCMAKE\_BUILD\_TYPE=Release -DCONFIG\_MULTITHREAD=1 -DCONFIG\_PIC=1 -DCONFIG\_REALTIME\_ONLY=1 -DCONFIG\_RUNTIME\_CPU\_DETECT=1 -DCONFIG\_WEBM\_IO=0} \\
        \hline
        AV1      & SVT-AV1 0.8.4  & \url{https://github.com/OpenVisualCloud/SVT-AV1}    & \texttt{build.sh release} \\
        \hline
        VP8, VP9 & libvpx 1.9.0   & \url{https://chromium.googlesource.com/webm/libvpx} & \texttt{configure --enable-pic --enable-realtime-only --enable-multi-res-encoding --disable-debug --cpu=x86-64} \\
        \hline
        H.265    & x265 release 3.5       & \url{https://github.com/videolan/x265}      & \texttt{cmake -DCMAKE\_BUILD\_TYPE=Release} \\
        \hline
        H.264    & x264 stable \texttt{cde9a933} & \url{https://code.videolan.org/videolan/x264} & \texttt{make} \\
        \hline
        H.264    & openh264 2.1.1 & \url{https://github.com/cisco/openh264}             & \texttt{make OS=linux ARCH=x86\_64}
    \end{tabular}
    \caption{Video encoders used in this study}
    \label{tab:codecs}
\end{table*}

\subsection{Video Codecs}

We will compare the performance of eight encoders, namely aomenc (default and real-time settings) and SvtAv1EncApp for AV1, vpxenc for VP8 and VP9, x265 for HEVC, x264 and h264enc for H.264, compiled in their real-time mode when available, and using speed~8 when applicable, using various bitrate targets.

Table~\ref{tab:codecs} gives for each codec all the information needed to reproduce the results: the version we have used, where to get their source code and which options we selected to compile them.

To give an insight of encoding performance difference between real-time mode and non real-time mode, only for AV1, we have compiled a second version of AOM encoder aomenc compiled using the option \texttt{-DCONFIG\_REALTIME\_ONLY=0}. We have selected speed option \texttt{--cpu-used=3}.

For the real-time version of aomenc, we have selected the highest speed option \texttt{--cpu-used=8}. We will refer to aomenc in real-time mode as aomenc-rt (encoder called with option \texttt{--rt}, and to aomenc not in real-time mode as aomenc-good (encoder called with good quality option \texttt{--good}).

We also selected the highest speed option \texttt{--preset 8} for SVT-AV1.

We selected the setting \texttt{--preset superfast} for x265, and the setting \texttt{--preset medium} for x264. In both case, it was enough to reach the max 50~fps target.

The options used at run time to launch each codec are given in Table~\ref{tab:encoders_options}.

Compilation of encoders and encoding of videos have been performed on a Dell{\texttrademark} OptiPlex 5050 with processor Intel{\textregistered} Core{\texttrademark} i7-7700T 8~cores at 2.90 GHz and 16~GB memory running Ubuntu Desktop 20.04.1 64~bits operating system.

\begin{table*}[tb]
    \centering
    \begin{tabular}{c|p{16cm}}
        Codec & Encoder command options \\
        \hline
        aomenc rt   & \texttt{aomenc --codec=av1 --profile=0 --kf-max-dist=90000 --end-usage=cbr --min-q=1 --max-q=63 --undershoot-pct=50 --overshoot-pct=50 --buf-sz=1000 --buf-initial-sz=500 --buf-optimal-sz=600 --max-intra-rate=300 --passes=1 --rt --lag-in-frames=0 --error-resilient=0 --tile-columns=0 --aq-mode=3 --enable-obmc=0 --enable-global-motion=0 --enable-warped-motion=0 --deltaq-mode=0 --enable-tpl-model=0 --mode-cost-upd-freq=2 --coeff-cost-upd-freq=2 --enable-ref-frame-mvs=0 --mv-cost-upd-freq=3 --enable-order-hint=0 --cpu-used=8 --threads=8 --end-usage=cbr --target-bitrate=xxx --fps=25/1} \\
        \hline
        aomenc good & \texttt{aomenc --codec=av1 --good --passes=1 --cpu-used=3 --threads=8 --lag-in-frames=25 --min-q=0 --max-q=63 --auto-alt-ref=1 --kf-max-dist=150 --kf-min-dist=0 --drop-frame=0 --static-thresh=0 --arnr-maxframes=7 --arnr-strength=5 --sharpness=0 --undershoot-pct=100 --overshoot-pct=100 --frame-parallel=0 --tile-columns=0 --profile=0 --target-bitrate=xxx --fps=25/1} \\
        \hline
        SVT-AV1 & \texttt{SvtAv1EncApp --tbr xxx --fps 25 --preset 8 --pred-struct 0 --profile 0 --rc 2 --min-qp 1 --max-qp 63 --vbv-bufsize 1 --tile-columns 0 --enable-global-motion 1 --enable-local-warp 0 --adaptive-quantization 0} \\
        \hline
        VP8      & \texttt{vpxenc --codec=vp8 --lag-in-frames=0 --error-resilient=0 --kf-max-dist=90000 --static-thresh=0 --end-usage=cbr --undershoot-pct=50 --overshoot-pct=50 --buf-sz=1000 --buf-initial-sz=500 --buf-optimal-sz=600 --max-intra-rate=300 --resize-allowed=0 --drop-frame=0 --passes=1 --rt --noise-sensitivity=0 --cpu-used=-6 --threads=8 --min-q=1 --max-q=63 --screen-content-mode=0 --target-bitrate=xxx --fps=25/1} \\
        \hline
        VP9      & \texttt{vpxenc --codec=vp9 --lag-in-frames=0 --error-resilient=0 --kf-max-dist=90000 --static-thresh=0 --end-usage=cbr --undershoot-pct=50 --overshoot-pct=50 --buf-sz=1000 --buf-initial-sz=500 --buf-optimal-sz=600 --max-intra-rate=300 --resize-allowed=0 --drop-frame=0 --passes=1 --rt --noise-sensitivity=0 --cpu-used=7 --threads=8 --profile=0 --min-q=1 --max-q=63 --tile-columns=0 --aq-mode=3 --target-bitrate=xxx --fps=25/1} \\
        \hline
        x265     & \texttt{x265 --frame-threads 0 --preset superfast --bitrate xxx --fps 25} \\
        \hline
        x264     & \texttt{x264 --preset medium --bitrate xxx --fps 25 --demuxer raw} \\
        \hline
        openh264 & \texttt{h264enc -rc 0 -complexity 2 -denois 0 -scene 0 -bgd 0 -fs 0 -numl 1 -tarb 0 xxx -frout 0 25}
    \end{tabular}
    \caption{Options used for encoders at run time (encoding has been performed at ten target bitrates: 800, 900, 1000, 1250, 1500, 1750, 2000, 2500, 5000 and 10000 kbps)}
    \label{tab:encoders_options}
\end{table*}

\begin{figure}[tb]
    \centering
    \includegraphics[width=0.5\textwidth]{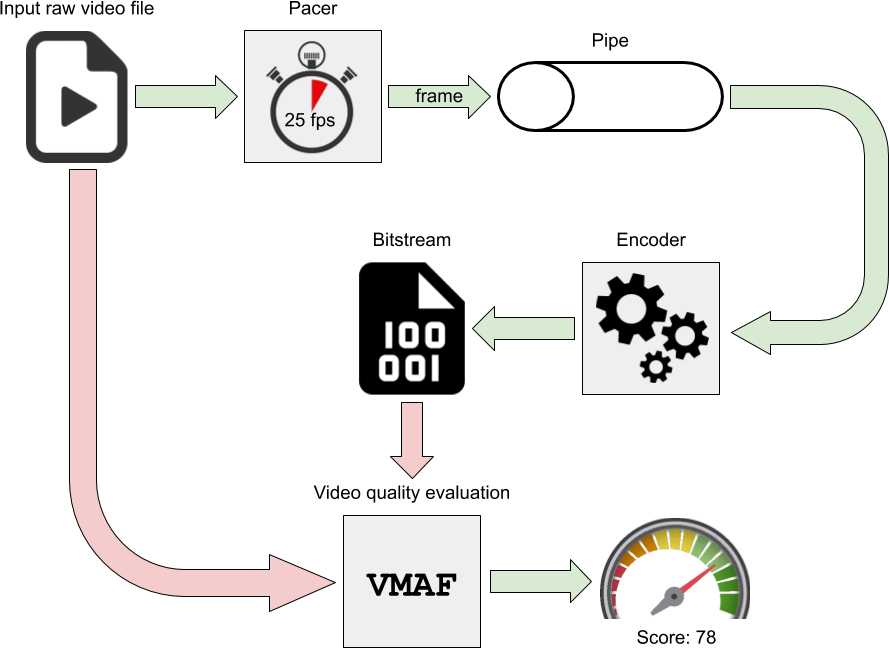}
    \caption{Evaluation of video encoder in real-time mode}
    \label{fig:pacer}
\end{figure}

\subsection{Real-Time Encoding Evaluation Process}

The evaluation of a video encoder is usually performed by letting the encoder to read a video file to be encoded. This is not a realistic mode of operation for real-time where the frames to be encoded are only available after some delay. For example, when encoding frames received from a camera at a rate of 25~fps, each new frame is available only after a delay of 1/25 second, that is 40~ms. Even if the encoder is able to encode a frame in 5~ms, it has to wait until the next frame becomes available. We are \emph{starving} the encoder. 

To evaluate the encoders in a more realistic process, we introduce a pacer program between the video file and the encoder (see Fig.~\ref{fig:pacer}). The objective of the pacer is to deliver frames to the encoder only at a selected frame rate.

A raw video file to be encoded is read by the pacer program. The pacer is in charge of simulating the real-time delivery of images at a selected rate. When the pacer reads frames from a 25~fps video file, it outputs a frame every 1/25th second. For a 50~fps video file, the pacer will output a frame every 1/50th second. Frames delivered by the pacer at the selected rate are written to a Unix pipe. The video encoder being tested reads frames from the Unix pipe. The output of the encoder is a bitstream corresponding to the encoded video file. At last, we evaluate the quality of the encoded video by using VMAF video quality assessment tool.




\begin{table}[tb]
    \centering
    \begin{tabular}{c|c|c|c|c|c|c}
        Video  & openh264 & x264     & VP8      & VP9      & x265    & SVT \\
        \hline
        BS25   & $-61.43$ & $-12.70$ & $-39.29$ &  $-9.49$ &  $7.63$ &  $3.61$ \\
        PA25   & $-56.12$ & $-22.39$ & $-40.35$ & $-10.91$ & $16.10$ & $28.94$ \\
        RB25   & $-52.39$ & $-42.83$ & $-32.72$ &  $-4.03$ & $-0.97$ & $17.38$ \\
        RH25   & $-50.78$ & $-26.08$ & $-31.52$ &  $-6.93$ & $16.22$ & $18.74$ \\
        ST25   & $-47.44$ &  $-6.58$ & $-33.41$ & $-16.50$ & $40.99$ & $33.33$ \\
        SF25   & $-65.01$ &  $-2.26$ & $-26.04$ & $-15.69$ & $34.49$ & $34.65$ \\
        TR25   & $-68.18$ & $-15.59$ & $-25.96$ &  $-1.16$ & $20.84$ & $31.39$ \\
        \hline
        Avg 25 & $-57.34$ & $-18.35$ & $-32.76$ &  $-9.24$ & $19.33$ & $24.01$ \\
        \hline
        CR50   & $-61.73$ & $-24.24$ & $-41.74$ & $-13.63$ & $-1.65$ &  $8.90$ \\
        DT50   & $-66.20$ & $-21.68$ & $-34.05$ & $-15.33$ & $22.54$ & $39.03$ \\
        IT50   & $-68.62$ & $-20.14$ & $-40.96$ &  $-8.31$ &  $7.52$ & $19.98$ \\
        OT50   & $-62.80$ & $-18.55$ & $-56.18$ & $-14.95$ & $12.56$ & $22.24$ \\
        PJ50   & $-62.98$ &  $-6.03$ & $-34.54$ & $-13.78$ & $14.10$ & $39.18$ \\
        \hline
        Avg 50 & $-64.47$ & $-18.13$ & $-41.49$ & $-13.20$ & $11.01$ & $25.87$
    \end{tabular}
    \caption{BD-rate for aomenc-rt8 (Note: SVT = SVT-AV1)}
    \label{tab:bdrate}
\end{table}

\begin{table}[tb]
    \centering
    \begin{tabular}{c|c|c|c|c|c|c}
        Video  & openh264 & x264    & VP8     & VP9    & x265    & SVT \\
        \hline
        BS25   & $12.61$  &  $1.36$ &  $3.90$ & $0.96$ & $-0.55$ & $-0.31$ \\
        PA25   & $14.70$  &  $4.41$ &  $6.62$ & $1.76$ & $-2.09$ & $-3.39$ \\
        RB25   & $24.88$  & $14.91$ & $13.17$ & $1.21$ &  $0.35$ & $-4.36$ \\
        RH25   & $10.59$  &  $4.51$ &  $4.30$ & $0.82$ & $-1.64$ & $-1.84$ \\
        ST25   &  $8.63$  &  $0.61$ &  $3.98$ & $1.14$ & $-1.79$ & $-1.76$ \\
        SF25   & $12.11$  &  $0.34$ &  $2.88$ & $1.10$ & $-1.54$ & $-1.52$ \\
        TR25   & $23.35$  &  $3.75$ &  $4.14$ & $0.14$ & $-3.20$ & $-4.62$ \\
        \hline
        Avg 25 & $15.27$  &  $4.27$ &  $5.57$ & $1.02$ & $-1.49$ & $-2.54$ \\
        \hline
        CR50   & $24.60$  &  $8.06$ & $12.78$ & $3.94$ &  $0.44$ & $-2.06$ \\
        DT50   & $25.06$  &  $5.81$ &  $9.99$ & $3.72$ & $-4.27$ & $-6.82$ \\
        IT50   & $21.77$  &  $5.02$ &  $4.85$ & $1.44$ & $-1.21$ & $-2.90$ \\
        OT50   & $14.64$  &  $3.03$ &  $7.11$ & $1.47$ & $-1.03$ & $-1.67$ \\
        PJ50   & $21.70$  &  $1.75$ &  $8.33$ & $3.51$ & $-2.81$ & $-7.17$ \\
        \hline
        Avg 50 & $21.55$  &  $4.73$ &  $8.61$ & $2.82$ & $-1.78$ & $-4.12$
    \end{tabular}
    \caption{BD-VMAF for aomenc-rt8 (Note: SVT = SVT-AV1)}
    \label{tab:bdvmaf}
\end{table}

\section{Results and Analysis}

We measured the VMAF scores for each encoder at ten different target bitrates: 800, 900, 1000, 1250, 1500, 1750, 2000, 2500, 5000 and 10000 kbps (see VMAF graphs in Figures \ref{fig:vmaf_bs} to \ref{fig:vmaf_pj}), and computed the BD-rates from the VMAF curves according to bitrate (see Tables \ref{tab:bdrate} and \ref{tab:bdvmaf}).

\subsection{ Pacer: Latency Impact of Real-Time Streams }

\begin{figure}[h]
    \centering
    \begin{subfigure}[b]{0.24\textwidth}
        \centering
        \includegraphics[width=\textwidth]{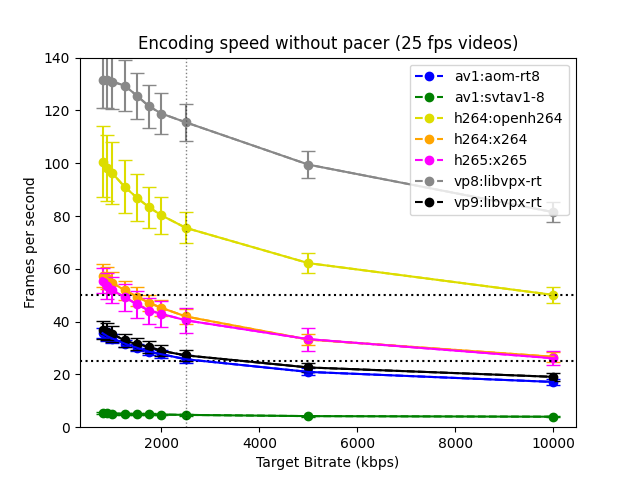}
        \caption{Average encoding speed of the seven 25 fps videos without pacer}
        \label{fig:encoding-speed25}
    \end{subfigure}%
    ~
    \begin{subfigure}[b]{0.24\textwidth}
        \centering
        \includegraphics[width=\textwidth]{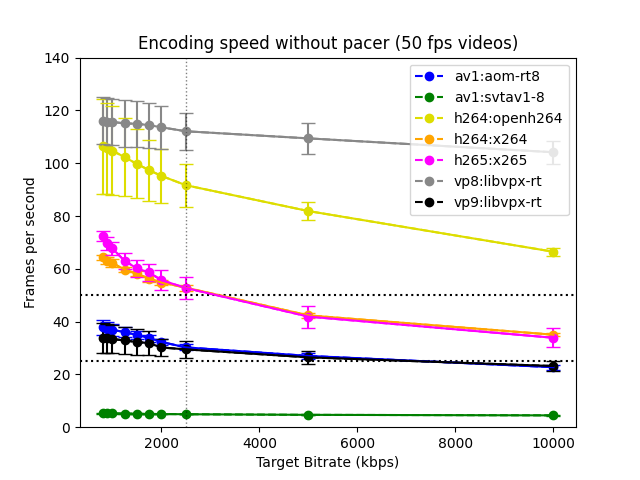}
        \caption{Average encoding speed of the five 50 fps videos without pacer}
        \label{fig:encoding-speed50}
    \end{subfigure}
    \newline
    \begin{subfigure}[b]{0.24\textwidth}
        \centering
        \includegraphics[width=\textwidth]{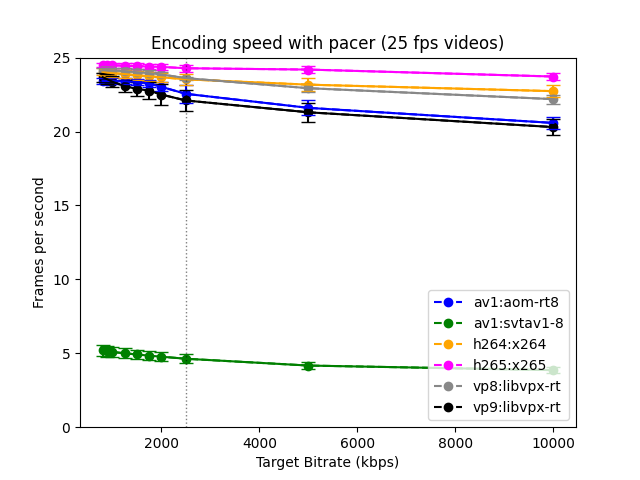}
        \caption{Average encoding speed of the seven 25 fps videos with pacer}
        \label{fig:encoding-speed_pacer25}
    \end{subfigure}%
    ~
    \begin{subfigure}[b]{0.24\textwidth}
        \centering
        \includegraphics[width=\textwidth]{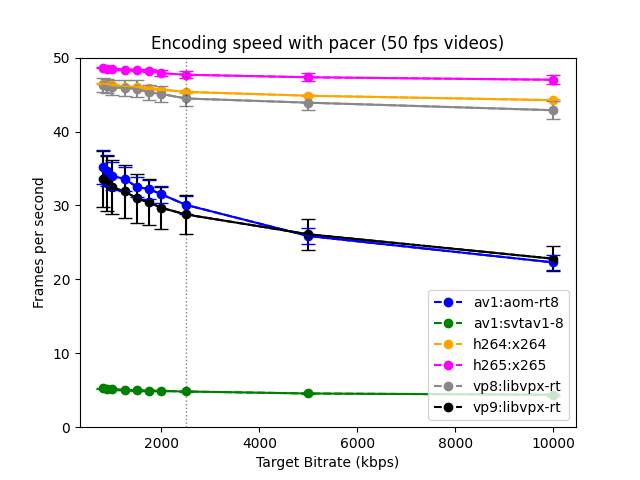}
        \caption{Average encoding speed of the five 50 fps videos with pacer}
        \label{fig:encoding-speed_pcaer50}
    \end{subfigure}
    \caption{Average encoding speed of 1080p videos on a Dell{\texttrademark} OptiPlex 5050 with processor Intel{\textregistered} Core{\texttrademark} i7-7700T 8~cores at 2.90~GHz and 16~GB memory running Ubuntu Desktop 20.04.1 64~bits operating system.}
    \label{fig:encoding-speed}
\end{figure}

We measured the encoding throughput first without using the pacer (the encoder reads the frames directly from the video file), which, supposing the encoding speed is limited by the encoder but not by the I/O speed, provides the maximum speed at which the encoder can operate given the input content, the bitrate target, and the hardware.
We then measured the throughput using the pacer. 

Fig.~\ref{fig:encoding-speed25} and \ref{fig:encoding-speed50} show the encoding throughput achieved by the encoders when they are not limited by the pacer. At the threshold bitrate of 2500~kbps or less, all the encoders are able to deliver 25~fps or more on our test computer. except SVT-AV1 which at best reach only 5.3~fps.

It is interesting to note that, when requested to encode for a~rate of 50~fps instead of 25~fps, all the encoders perform their compression faster, although three of the five videos in the 50~fps group are rather difficult to be encoded. This increase in compression speed allows x264 and x265 to reach 50~fps at the threshold bitrate of 2500~kbps, while VP9 and aomenc-rt are not able to deliver more than 30~fps.

Generally, the encoders work at a more homogeneous speed at higher bitrates as shown by the shorter standard deviation bars, while the measured frame rate varies much more for lower bitrates. In fact, at low bitrates, the encoders process videos easy to encode (Blue-sky, Pedestrian-area, Rush-hour, Station2, Sunflower, Tractor, In-to-tree, Old-town0cross) much faster than videos harder to encode (Riverbed, Crowd-run, Ducks-take-off, Park-run).

Apart from the settings, this is the conditions under which most codec studies are being done, and performances reported.

One can see that in the case of x264 and x265, the latency is the bottleneck of the throughput and not the encoding speed as it can encode 50~fps media content faster than 25~fps. Supposing a constant I/O speed, reading a frame (latency) is constant whether the media content is originally captured at 25 or 50~fps, however, the complexity (linear to the number of pixels to encode) per second is twice as much.

This is the opposite for libvpx, for which we can clearly see that encoding 50~fps content is slower than encoding 25~fps whether with VP8 or VP9.

Introducing the pacer, we measured the frame rates as shown in Fig.~\ref{fig:encoding-speed_pacer25} and \ref{fig:encoding-speed_pcaer50}. 

All the encoders which could deliver more than 25 fps in the previous experiment have no problem delivering with a paced input. SVT-AV1 remains way too slow for real-time.

All the encoders which could deliver more than 50fps in the previous experiment, like VP8 and x264 here again have no problem. Interestingly x265 is performing extremely well. It is also interesting to note that aomenc-rt is slightly faster than VP9, while one would expect the opposite. Right now we can conclude that there is no CPU penalty when moving from using VP9 to using AV1 real-time.

\subsection{Interpretation of BD-rate and BD-VMAF}

A BD-rate is a measure of the average percentage bitrate savings that can be obtained for the same visual quality level. This measure is computed over the range of quality levels that are common to two curves.

For example, let consider VMAF scores on Fig.~\ref{fig:vmaf_du} for Ducks-take-off (DT50) video.
We want to compute the bitrate savings at same VMAF level of aomenc-rt8 (blue curve, VMAF range from 27 to 61) as compared to x264 (orange curve, VMAF range from 6 to 55). The common VMAF range for these two curves is 27 to 55.
Using that common quality range, we compute the average bitrate savings by calculating the area between the curves (to the left of the x264 orange curve and to the right of the aomenc-rt8 blue curve), and we divide it by the area to the left of the aomenc-rt8 blue curve up to the right of the y-axis.
We get a BD-rate of $-21.68$ as shown in Table~\ref{tab:bdrate}. It is a negative value, which means that there is a reduction in bitrate for aomenc-rt8 as compared to x264. The interpretation of this value is that for the same VMAF score, we may expect that aomenc-rt8 gives in average a 21.68\% bitrate savings as compared to x264.

Similarly, we can compute the average visual quality improvement for the same bitrate between aomenc-rt8 and x264 by switching the variables.
Using the same example, we look for the common bitrate range between aomenc-rt8 (blue curve, bitrate range from 2160 to 9925 kbps) and x264 (orange curve, bitrate range from 810 to 10000 kbps). The common bitrate range for these two curves is 2160 to 9925 kbps.
Using that common bitrate range, we compute the average VMAF improvement by calculating the area between the curves (to the bottom of aomenc-rt8 blue curve and to the top of x264 orange curve), and we divide it by the area to the bottom of the x264 orange curve down to the top of the x-axis.
We get a BD-VMAF of $5.81$ as shown in Table~\ref{tab:bdvmaf}. It is a positive value, which means that there is an increase in VMAF score for aomenc-rt8 as compared to x264. The interpretation of this value is that for the same bitrate, we may expect that aomenc-rt8 gives in average a VMAF score $5.81$ points higher than x264.

\subsection{Discussion}

\begin{figure*}[p]
    \centering
    \begin{subfigure}[b]{0.33\textwidth}
        \centering
        \includegraphics[width=\textwidth]{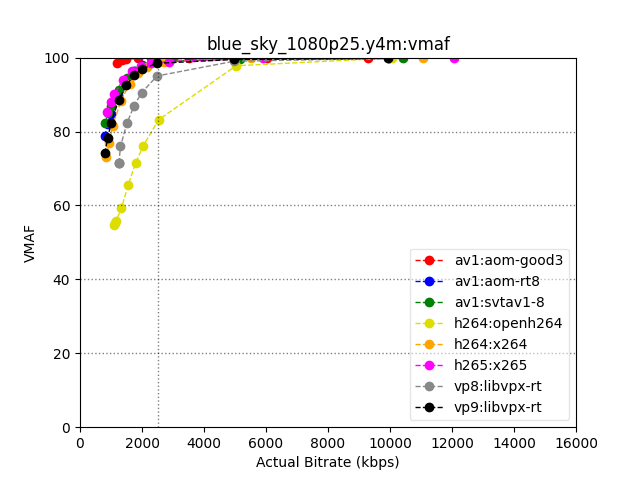}
        \caption{Blue sky (25 fps) BS25}
        \label{fig:vmaf_bs}
    \end{subfigure}%
    ~
    \begin{subfigure}[b]{0.33\textwidth}
        \centering
        \includegraphics[width=\textwidth]{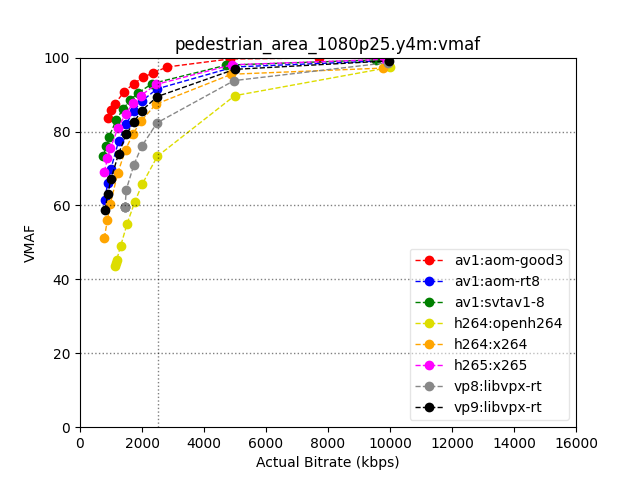}
        \caption{Pedestrian area (25 fps) PA25}
        \label{fig:vmaf_pa}
    \end{subfigure}%
    ~
    \begin{subfigure}[b]{0.33\textwidth}
        \centering
        \includegraphics[width=\textwidth]{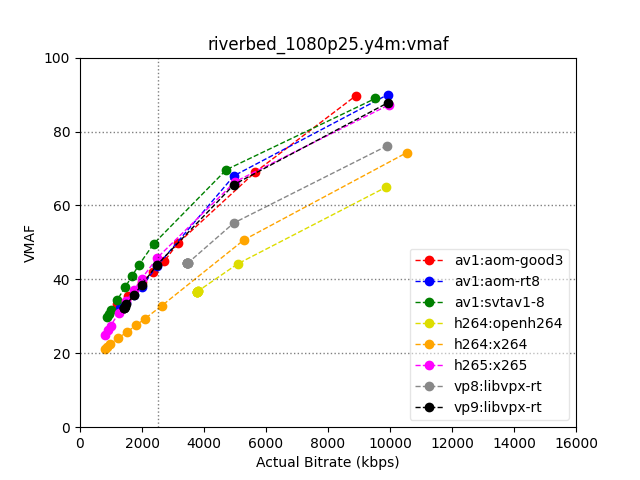}
        \caption{Riverbed (25 fps) RB25}
        \label{fig:vmaf_rb}
    \end{subfigure}
    \newline
    \begin{subfigure}[b]{0.33\textwidth}
        \centering
        \includegraphics[width=\textwidth]{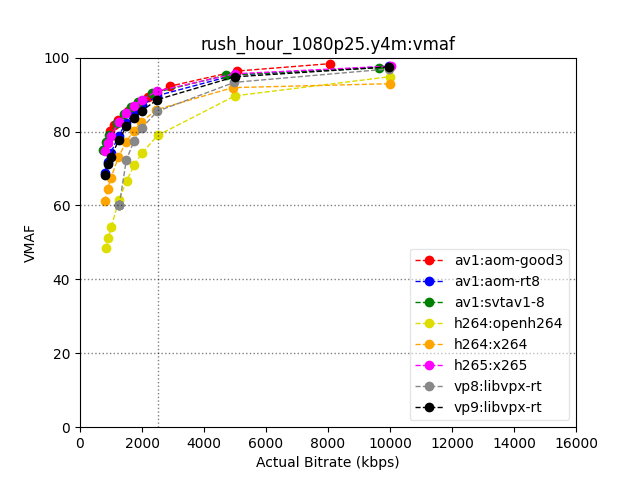}
        \caption{Rush hour (25 fps) RH25}
        \label{fig:vmaf_rh}
    \end{subfigure}%
    ~
    \begin{subfigure}[b]{0.33\textwidth}
        \centering
        \includegraphics[width=\textwidth]{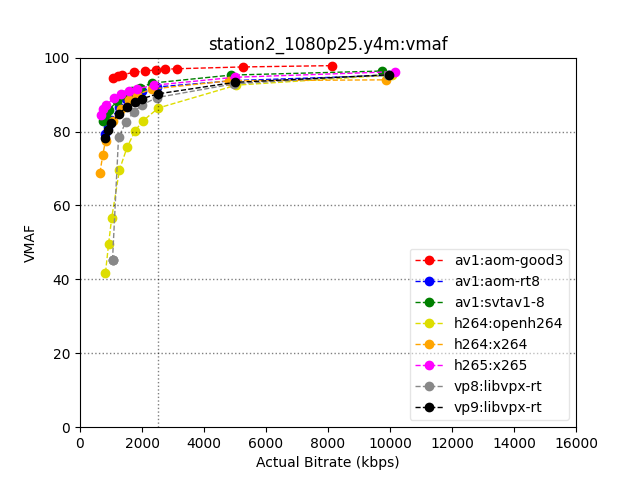}
        \caption{Station2 (25 fps) ST25}
        \label{fig:vmaf_st}
    \end{subfigure}%
    ~
    \begin{subfigure}[b]{0.33\textwidth}
        \centering
        \includegraphics[width=\textwidth]{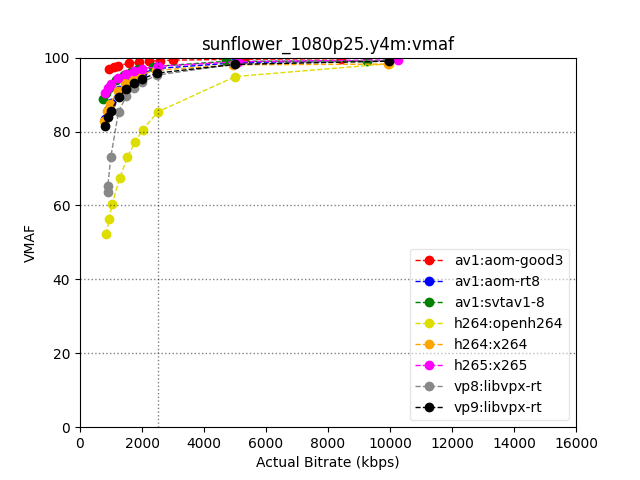}
        \caption{Sunflower (25 fps) SF25}
        \label{fig:vmaf_sf}
    \end{subfigure}
    \newline
    \begin{subfigure}[b]{0.33\textwidth}
        \centering
        \includegraphics[width=\textwidth]{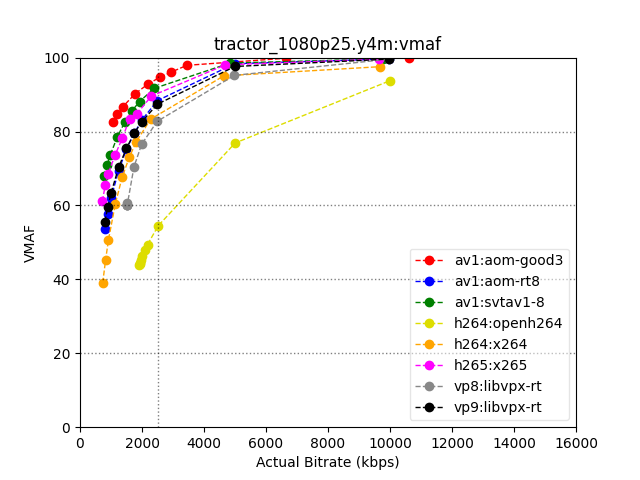}
        \caption{Tractor (25 fps) TR25}
        \label{fig:vmaf_tr}
    \end{subfigure}%
    ~
    \begin{subfigure}[b]{0.33\textwidth}
        \centering
        \includegraphics[width=\textwidth]{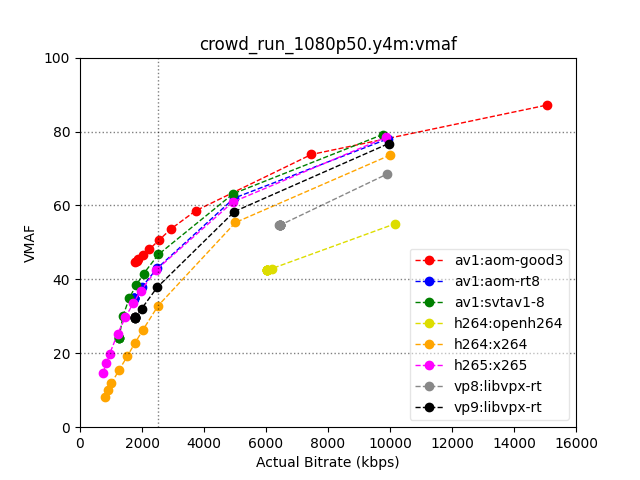}
        \caption{Crowd run (50 fps) CR50}
        \label{fig:vmaf_cr}
    \end{subfigure}%
    ~
    \begin{subfigure}[b]{0.33\textwidth}
        \centering
        \includegraphics[width=\textwidth]{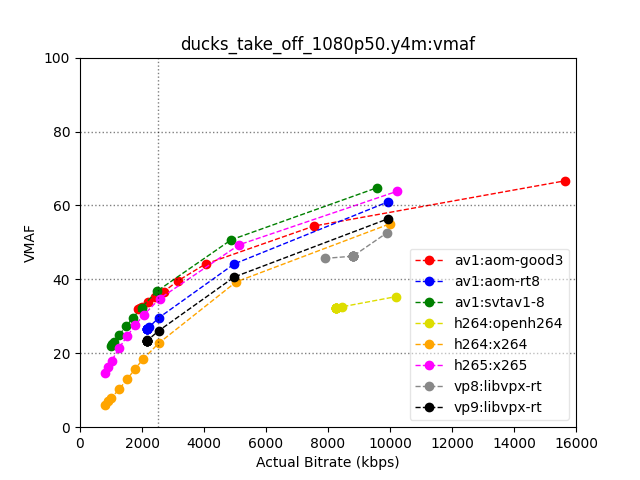}
        \caption{Ducks take off (50 fps) DT50}
        \label{fig:vmaf_du}
    \end{subfigure}
    \newline
    \begin{subfigure}[b]{0.33\textwidth}
        \centering
        \includegraphics[width=\textwidth]{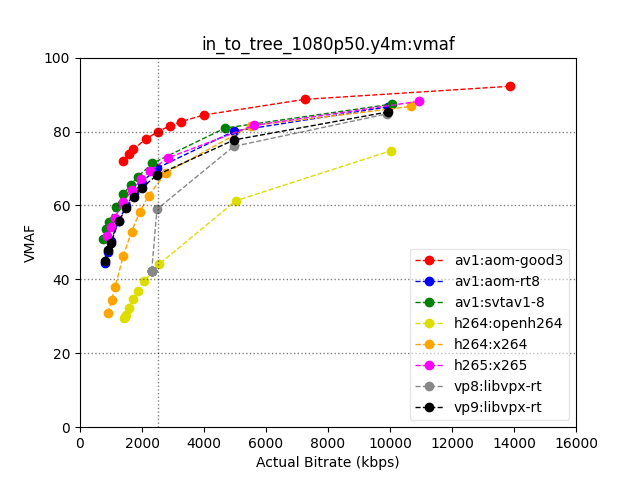}
        \caption{In to tree (50 fps) IT50}
        \label{fig:vmaf_it}
    \end{subfigure}%
    ~
    \begin{subfigure}[b]{0.33\textwidth}
        \centering
        \includegraphics[width=\textwidth]{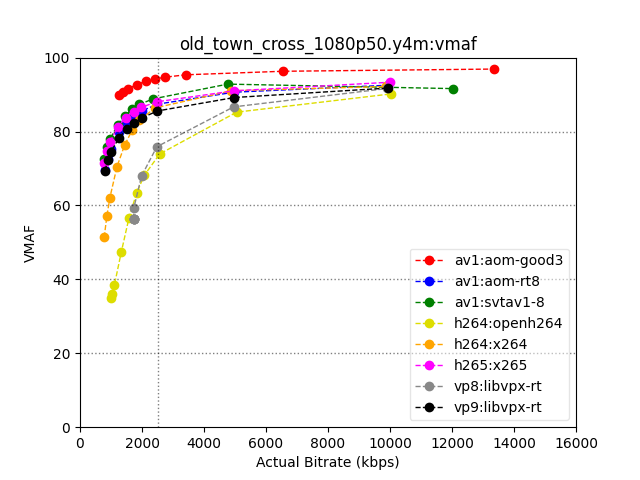}
        \caption{Old town cross (50 fps) OT50}
        \label{fig:vmaf_ot}
    \end{subfigure}%
    ~
    \begin{subfigure}[b]{0.33\textwidth}
        \centering
        \includegraphics[width=\textwidth]{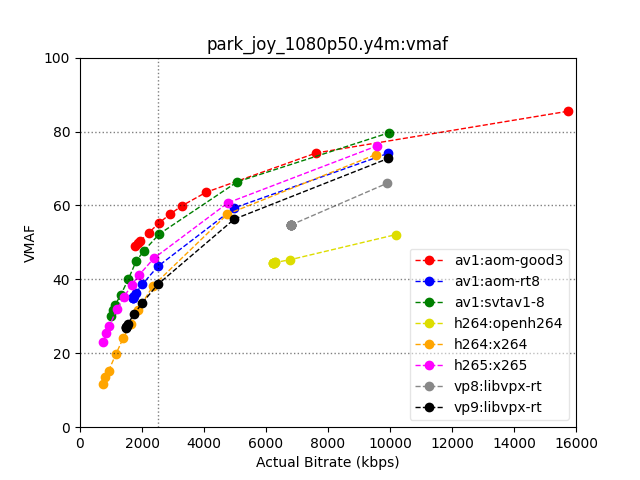}
        \caption{Park joy (50 fps) PJ50}
        \label{fig:vmaf_pj}
    \end{subfigure}%
    \caption{VMAF scores according to bitrate}
    \label{fig:vmaf_scores}
\end{figure*}

VMAF scores are reported in Fig.~\ref{fig:vmaf_bs} to \ref{fig:vmaf_pj}. We have highlighted a threshold of 2500~kbps for bitrate as this value is known to be a hard-coded maximum for WebRTC within Chromium browser, although native applications using WebRTC are not concerned by such a limitation.

\subsubsection{25~fps datasets}

VMAF scores show that all the videos of the 25~fps group except Riverbed are relatively easy to encode. The curves are close or very close to the perfect score of 100 for most bitrates. It is only at low bitrate of roughly 1000~kbps or lower that VMAF score gets lower than 80. The quality of videos encoded by openh264 is worsening at higher bitrates and faster than for the other encoders. VP8 is also showing a slightly lower quality of videos than the other encoders. At the threshold of 2500~kbps bitrate, the quality of encoded videos is excellent or good, above VMAF score of~80.

Riverbed is the only video of the 25~fps group being difficult to be encoded with good quality. Although this video is static, waves on surface of water and reflection of light on the waves are challenging for encoders.
Openh264 and VP8 are not able to encode this video at the threshold bitrate of 2500~kbps. The lowest bitrate delivered by openh264 for Riverbed is 3760~kbps, while it is 3460~kbps for VP8.
VMAF rating at the threshold bitrate is only between 30 and 50, which is poor quality.

\subsubsection{50~fps datasets}

The group of 50~fps videos was more difficult to encode with good quality than the group of 25~fps videos. Ducks-take-off was the most challenging video to be encoded. Interestingly, like Riverbed, it is a static video showing water with waves.

At the threshold bitrate of 2500~kbps, VMAf score is lower than 80 for the videos in the group of 50~fps, except for video Old-town-cross which has a very slow motion.

We notice again that openh264 provides noticeable lower VMAF scores than the other encoders.
There are three videos, Crowd-run, Ducks-take-off and Park-joy, which encoders openh264 and VP8 are unable to encode at the threshold bitrate of 2500~kbps. They would require a target bitrate of 6000~kbps or more.

The other encoders, aomenc-rt8, SVT-AV1, x265, x264 and VP9, are able to encode all the twelve videos with a rather similar quality. The quality of videos encoded by SVT-AV1 is always the best, x265 comes second and aomenc-rt8 is third.

From Table~\ref{tab:bdvmaf} giving BD-VMAF for aomenc-rt8, we can get the ranking of the encoders relatively to aomenc-rt8.
We can see that for the same bitrate, VMAF score of SVT-AV1 $>$ x265 $>$ aomenc-rt8 $>$ VP9 $>$ x264 $>$ VP8 $>$ openh264.

Among the two encoders studied for AV1, SVT-AV1 provides a much better coding efficiency than aomenc-rt real-time. However, SVT-AV1 is lacking an efficient real-time mode. It took SVT-AV1 about 7~times as long as aomenc-rt to encode the same video clips, although both were run using the same speed of~8.

\section{conclusion}

\subsection{Latency}

Pre-recorded content provide encoders with the capacity to fill up buffers to increase the coding efficiency without increasing the latency too much. The same buffers which are filled at I/O speed for pre-recorded content need to wait for frame to be acquired in live, real-time and interactive use case, making any operation that requires frame buffers prohibitive.

Very often, benchmarks are only provided for pre-recorded content, and cannot be directly translated into the real-time configuration. One of the contribution of this paper is a process to compute fair performance comparison of encoders in a real-time situation, while still using standard files as input. We~think it's going to help extend existing test beds to be able to better assess performances of all codecs in what has become a much more important use case.

\subsection{Coding Efficiency}

It is outside the scope of this paper, but it is interesting to note that SVT-AV1 at speed~8 has a coding efficiency similar or better than aomenc-good at speed~3.


The non-realtime version of aomenc has a better coding efficiency than the real-time version. The real-time version of AOM seems to be about 33\% less efficient. While theoretically interesting, this has but little practical interest, since the default aomenc encoder, even at its maximum setting of speed~6 will have too much latency to be used in interactive case. The interesting question is: what do I gain or lose when switching from one real-time codec to another, and which ones can achieve the lowest latency.

We find that the real-time modes of the codecs are generally performing relatively to each other as they would with their non real-time version, i.e AV1 better than VP9 better than VP8, and HEVC better than AVC. The exception is HEVC which has a better coding efficiency than aomenc-rt real-time, while in the non-real time mode, it is aomenc that is reported to have a better coding efficency than HEVC~\cite{chen2020}.

x265 exhibited excellent coding efficiency. It is able to encode 1080p videos in real-time at 50~fps. It looks like its speed settings has also very good (low) latency. The authors regret that the licensing situation of HEVC is still complicated.


We have shown that one can expect an average of 11\% less bandwidth usage for the same video quality with aomenc-rt than with VP9-rt, 37\% less than with VP8.

The x264 implementation of H.264 is more or less often in par with VP8 for quality, while the openh264 implementation of H.264 exhibits generally lower coding efficiency on the 12~video clips of this study. This illustrates that encoder implementations of the same codec can vary a lot in quality, and one should compare implementations and not codecs directly.

Codec implementations improve with time, so it is likely that the sitll young implementation of AV1 codecs will improve with time.

\section{Future Work}
This is the first attempt at comparing several real-time versions of encoders, so there is a lot of room for improvement.

One obvious way to improve the results is to test on more datasets, which could e.g. include different types of content (high-motion videos, animation, live video games), different resolutions (480p a.k.a. DVD size, 4k, 8k).
We think there is still a lot of work to be done to directly measure latency of encoders.

There are many more implementation of codecs out there, and testing different implementations, including hardware implementation would add significant value for readers.

\section*{Acknowledgement}
The authors would like to thank Andrew Johnson, CTO of Nira Inc., Ioannis Katsavounidis from Netflix, and Google collaborators (Marco Paniconi, Fyodor Kyslov, Michael Horowitz,
Yaowu Xu, Jerome Jiang, Danil Chapovalov, ...) for their valuable help, suggestions and comments.

\bibliographystyle{IEEEtran}
\bibliography{IEEEabrv,references.bib}

\end{document}